\documentclass[reprint,aps,amsmath,amssymb,nofootinbib,twoside,prd,showkeys,superscriptaddress]{revtex4-1}
\pdfoutput=1
\usepackage{verbatim}
\usepackage{comment}
\usepackage{graphicx}
\usepackage[T1]{fontenc}
\usepackage{epsfig}
\usepackage{bm}
\usepackage{tensor}
\usepackage{amssymb}
\usepackage{float}
\usepackage{amsmath}
\usepackage{subfigure}
\usepackage{dcolumn}
\usepackage[utf8]{inputenc}
\usepackage{cancel}
\usepackage[colorlinks]{hyperref}
\usepackage[usenames,dvipsnames]{color}
\hypersetup{
     breaklinks=true,
    pdfstartview={FitH},    
    colorlinks=true,       
    linkcolor=blue,          
    citecolor=red,        
    filecolor=magenta,      
    urlcolor=blue,           
    anchorcolor=green,      
    linktocpage=true
}

\def\doi{http://doi.org}

\newcommand{\aj}{Astron. J.}   
\newcommand{\aapr}{Astron. Astrophys. Rev.}   

\newcommand{\HCd}{\mathcal{H}}

\def\HCdt0{\tilde{\HCd}_{0}}

\usepackage{multirow}
\usepackage{amssymb}	
\usepackage{graphicx}
\usepackage{tikz}

\newcommand{\afffias}{Leibniz-Institut fur Astrophysik Potsdam, An der Sternwarte 16, D-14482 Potsdam, Germany}
\newcommand{\affhel}{Department of Physics, University of Helsinki, P.O.
Box 64, FI-00014 Helsinki, Finland}
\newcommand{\affhip}{Helsinki Institute of Physics, P.O. Box 64, FI-00014 University of Helsinki, Finland}

\begin{document}
\title{Galaxy groups in the presence of Cosmological Constant: \\ Increasing the Masses of Groups}
\author{David Benisty}
\email{benidav@aip.de}
\affiliation{\affhip}\affiliation{\afffias}
\author{Moshe M. Chaichian}
\email{masud.chaichian@helsinki.fi}
\affiliation{\affhip}\affiliation{\affhel}
\author{Anca Tureanu}
\email{anca.tureanu@helsinki.fi}
\affiliation{\affhip}\affiliation{\affhel}
\begin{abstract}
The boundaries of galaxy groups and clusters are defined by the interplay between the Newtonian attractive force and the decoupling from the local expansion of the Universe. This work extends the definition of a zero radial acceleration surface (ZRAS) and the turnaround surface (TS) for a general distribution of the masses in an expanding background, governed by the cosmological constant. We apply these definitions to different galaxy groups in the local Universe, mapping these groups up to ten megaparsec distances. We discuss the dipole and the quadrupole rate for the Local Group of Galaxies and the implementations on the Hubble diagram correction and galaxy groups virialization. With these definitions, we present the surfaces showing the interplay between the local expansion vs the local Newtonian attraction for galaxy groups in the local Universe. Further, we estimate the masses of different galaxy groups and show that the inclusion of the Cosmological Constant in the analysis predicts these masses to be higher by 5-10\%. For instance the Local Group of Galaxies is estimated to be $\left( 2.47 \pm 0.08\right) \cdot 10^{12} M_{\odot}$. For the groups with enough tracers close to the TS, the contribution of the Cosmological Constant makes the masses to be even higher. The results show the importance of including the local cosmic expansion in analyzing the Cosmic Flow of the local Universe.
\end{abstract}
\keywords{Dark Energy; Dark matter; Galaxy groups; Local Universe; Virial Theorem; Cosmic Flow}
\maketitle

\section{Introduction}
\label{sec:intro}
Dark energy is an unknown type of energy that influences the Universe on its largest scales~\cite{Peebles:2002gy}. Its primary impact is to propel the accelerating expansion of the Universe. The first observational evidence for dark energy emerged from supernova measurements~\cite{Pan-STARRS1:2017jku} and later was confirmed by the Cosmic Microwave Background radiation measurements~\cite{Planck:2018vyg} and the Baryon Acoustic Oscillations. Beyond its impact on large scales, dark energy exerts a significant influence on the Local Universe~\cite{Courtois:2013yfa}. Galaxy groups can be characterized by the radius of decoupling from cosmic expansion~\cite{2015AJ....149...54T}. In an isolated system, the dark energy creates a local repulsion force resisting the Newtonian attraction:
\begin{equation}
\ddot{r} = - G\frac{M}{r^2} + \frac{\Lambda c^2}{3}r \;,
\label{eq:eom}
\end{equation}
where $r$ is the separation, $M$ is the total mass of the binary system, $G$ is the Newtonian gravitational constant, $c$ is the speed of light and $\Lambda$ is the cosmological constant. It is possible to infer the cosmological constant from the cosmological parameters via: $\Lambda = 3 \Omega_\Lambda \left(H_0/c\right)^2\;$, where $H_0$ the Hubble parameter that estimates the current expansion rate, and $\Omega_\Lambda $ is the dark energy density rate.

Since the dark-energy contribution depends on the mass density, we define two useful surfaces: the Zero Radial Acceleration Surface (ZRAS) and the Turnaround Surface (TS). These two definitions are compatible for the spherical case, and in this paper we generalize them for a general density structure. 

\begin{itemize}
\item \textbf{Zero Radial Acceleration Surface (ZRAS)}, or zero gravity radial surface, is the surface where the total radial force on a point particle of the system is zero. For the spherical case, imposing $\ddot{r} = 0$ in Eq.~(\ref{eq:eom}) gives:
\begin{equation}
\bar{r}_0 \equiv \sqrt[3]{\frac{3 G M}{\Lambda c^2}} \approx 1.11\, \text{Mpc} \, \left(\frac{M}{10^{12}\,M_{\odot}}\right)^{1/3} \left(\frac{67 \, km/s/Mpc}{H_0}\right)^{2/3} ,
\label{eq:r0def}
\end{equation}
which is called the zero gravity radius. 
$M_{\odot}$ is the solar mass. There are different methods to measure the zero gravity radius through the local Hubble expansion~\cite{DelPopolo:2022sev} or by estimating the points at which the sphere breaks away from the general expansion and reaches a maximum radius~\cite{Pavlidou:2013zha,Faraoni:2015saa,Faraoni:2015zqa,Giusti:2019lha}. For the spherical case, the zero gravity radius is sufficient, but for a general density distribution one has to use the ZRAS, which reads~\cite{2003ARep...47..728D}:
 \begin{equation}
{\bold r} \cdot {\nabla} \Phi({\bf r}) = 0,
\label{eq:zasCon}
\end{equation}
where $\bold{r}$ is the position vector. As shown further on, this general definition separates the gravitational attraction domination areas from the expansion domination areas.

\item \textbf{Turnaround Surface (TS)} or \textbf{Zero Velocity Surface} -- In statistical mechanics, the virial theorem provides a general equation that relates the average over time of the total kinetic energy of a closed and stable system of discrete particles, bound by a conservative force, to the average of the potential energy of the system~\cite{Lahav:1991wc,Mota:2004pa,Wang:2005ad,2012GrCo...18....1C}. The conventional virial theorem can be modified by an extra term due to the contribution of the dark energy, besides the Newtonian attraction. By imposing that all the kinetic energy of the system is positive  ($K > 0$) one gets the condition~\cite{Nowakowski:2001zw}:
\begin{equation}
\frac{|U_G|}{2} < U_{\Lambda}.
\label{eq:con}
\end{equation}
For a point mass $M$ or a spherical density, the maximal radius of a test particle in a "virialized system"  ( i.e. a system which can be considered as a closed one and the virial theorem can be used for it) is $r < \bar{r}_0$. 

The condition for positive kinetic energy is extended for a general density case and gives the definition for the virialization surface. In the standard case where the dispersion of the velocities is known, the virial theorem gives the lowest possible radii for a closed system. In this case, it is possible to calculate the value of the average spherical radii. Consequently, there exists a surface that obeys the virial theorem for the particles inside the surface.  
\end{itemize}

Galaxy groups can be characterized by the radius of decoupling from cosmic expansion ~\cite{2015AJ....149...54T}. However, this definition has been made without the presence of dark energy. Ref.~\cite{2015AJ....149...54T,2015AJ....149..171T,2017ApJ...843...16K} consider the real space and ~\cite{Tempel:2016gdu,Tempel:2017dhe,Sorce:2018skr} consider the redshift space. In this {\it{letter}} we extend these definitions from a spherical symmetric case to a general density using multipole expansion and define the corresponding ZRAS and TS.

The structure of the paper is as follows: Section~\ref{sec:one} reviews the theoretical framework for a spherical density model with dark energy. Section~\ref{sec:twoGalaxies} extends the analysis for a galaxy pair. Section~\ref{sec:localUniverse} shows the applications for galaxies in the local Universe and maps the local galaxy complexes correspondingly. Section~\ref{sec:mul} generalizes the formulation for general density with multipole expansion. Section~\ref{sec:massesLambda} calculates the masses of certain groups of galaxies in the presence of the cosmological constant $\Lambda$. Section~\ref{sec:dis} contains the conclusion and summarizes the results of the work. 

\section{Point Particle and Spherical Symmetry}
\label{sec:one}

\paragraph{The space time} 
The de Sitter space is the simplest solution of Einstein's equation with a positive cosmological constant, and the Schwarzschild solution is the simplest spherically symmetric solution with a massive body in the origin. The de Sitter–Schwarzschild space-time is a combination of these two, and describes a massive body spherically centered in an otherwise de Sitter Universe. The potential of the de Sitter–Schwarzschild space-time reads: 
\begin{equation}
\Phi_{\text{SDS}}(r) = \frac{G M}{r} + \frac{\Lambda c^2}{6} r^2.
\label{eq_upointmass}
\end{equation}
The term captures the gravitational potential and dark energy contributions. It is thus a de~Sitter--Schwarzschild metric which reduces to the Schwarzschild metric in the limit of $\Lambda=0$.
 
\paragraph{Zero Acceleration Surface}  
From Eq.~(\ref{eq:zasCon}):
\begin{equation}
{\nabla} \Phi({\bf r}) = \left(\partial_r \Phi(r),0,0\right) = {0},
\end{equation}
one obtains the solution as in Eq.~(\ref{eq:r0def}). In this case, the radial acceleration is the same as the total acceleration.

\paragraph{Turnaround Surface}
The virial theorem provides a general equation that relates the average over time of the total kinetic energy of a stable system of discrete particles, bound by a conservative force, to that of the total potential energy of the system. The virial theorem relates the average of a potential energy with a power law of the form $U \sim r^{n}$, via the relation~\cite{Landau:1975pou,Goldstein2001}: $2 \langle K \rangle= n \langle U \rangle$, where $K$ is the kinetic energy of the system, $U$ is the potential energy of the system, and the averages are taken over a long time. The average yields:
\begin{equation}
2K + U_{G} = 2U_{\Lambda},
\label{eq:virGen}
\end{equation}
where $U_{G} $ is the gravitational potential energy and $U_{\Lambda} $ is the repulsion energy due to the cosmological constant. For point particles the terms read:
\begin{equation}
K = \frac{1}{2} \sum m_i v_i^2, \quad U_G = G \sum_{i<j} m_i m_j/r_{ij}, \quad 
U_\Lambda = \frac{\Lambda c^2}{6} \sum_{i} m_i r_i^2 . \label{eq:vir}
\end{equation}
For the local Universe galaxies, we assume that the dwarf galaxies are test particles around the giant ones~\cite{Karachentsev:2010nw}. Since the kinetic energy is positive, Eq.~(\ref{eq:virGen}) implies the condition~(\ref{eq:con}). Therefore, the kinetic energy of a virialized system is less in the presence of the dark energy background than in perfectly empty space. Dark energy partly cancels the matter attraction and as a result, the potential well of the system is not as deep as it would be in empty space.

As a simple model for the galaxy groups in the local Universe, it is assumed that most of the mass is concentrated in the massive galaxies, and the other dwarf galaxies are test particles with mass $m$. For test particles obeying the Eqs.~(\ref{eq:vir}) with the condition~(\ref{eq:con}), one finds $r < \bar{r}_0$, where $\bar{r}_0$ is the ZRAS for the massive galaxy $M$. It means that for a closed system within the zero radial acceleration surface. The estimate indicates that gravitation is stronger than repulsion within the volume
of a group, making the existence of gravitationally
bound systems with finite orbits possible. Based on this condition, it was suggested in \cite{Membrado:2016zjw} to use surface term effects as mass estimators. From this inequality, it is clear that this gives an upper value for the mass inside the spherical surface.

\section{Galaxy Groups with Axial Symmetry}
\label{sec:twoGalaxies}
\subsection{The surfaces}
\begin{figure}[t!]
\centering
\includegraphics[width=0.45\textwidth]{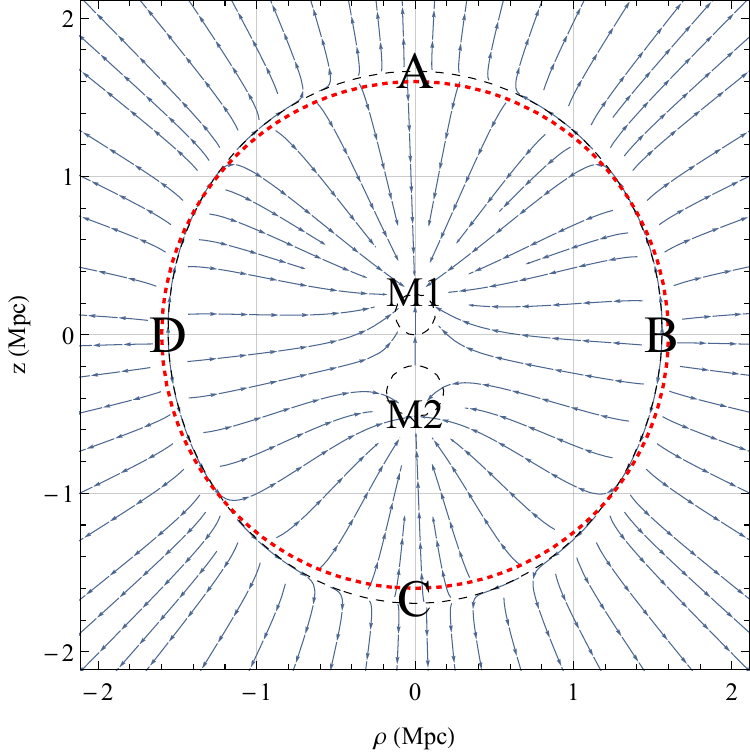} 
\caption{ \it{Stream lines for a galaxy pair with masses $M_{1,2}$ located at $\left(0,z_{1,2}\right)$. The stream lines shows the attractor for the gravitational force inside the surface described by the Eq.~(\ref{eq:zeroRadSur}), vs the repulsion field outside. The zero acceleration surface is presented in red line.}}
\label{fig:perDip}  
\end{figure}

\paragraph{The system} This section extends the definitions discussed above for galaxy pairs. For this model, most of the mass including the dark matter is located at the galaxies, and other members are considered to be point particles around the massive ones. The system has axial symmetry and, therefore, can be described with two coordinates, $(\rho,z)$, where $\rho^2 = x^2 + y^2$. The galaxies have masses $M_1$ and $M_2$, located at $(0,z_1)$ and $(0,z_2)$, respectively (where $z_2 < 0$). The center of mass of the system is located at the origin:
\begin{equation}
M_1 z_1 = - M_2 z_2.
\label{eq:CoM}
\end{equation}
The separation between the galaxies is denoted by $z_{12}$ and it obeys the relations: 
\begin{equation}
z_{1} = m_2 z_{12}, \quad z_{2} = -m_1 z_{12},
\end{equation}
where $m_{1,2}$ are the fractional masses from the total mass: $m_{1,2} = M_{1,2}/(M_1+M_2)$. Fig.~(\ref{fig:perDip}) shows an illustration of this system. A test particle around these galaxies feels the potential
\begin{equation}
\Phi(\rho,z) = \frac{G M_1}{\sqrt{\rho^2+(z-z_1)^2}}+\frac{G M_2}{\sqrt{\rho ^2+(z - z_2)^2}}+\frac{\Lambda c^2}{6} r^2,
\end{equation}
where $r^2 = \rho^2 + z^2$. There are two critical points where the total force is zero (points A and C) and  Fig.~(\ref{fig:perDip}) illustrates the critical points as well.

\paragraph{Critical points} The field acting on the test particle reads:
\begin{subequations}
\begin{equation}
F_z = \frac{\Lambda c^2 z}{3} - \frac{G M_1 (z-z_1)}{\left[\rho^2+(z-z_1)^2\right]^{3/2}} - \frac{G M_2 (z-z_2)}{\left[\rho ^2+(z-z_2)^2\right]^{3/2}},
\label{eq:fz}
\end{equation}
\begin{equation}
F_\rho = \frac{\Lambda c^2 \rho}{3}  - \frac{G M_1 \rho}{\left[\rho^2+(z-z_1)^2\right]^{3/2}}-\frac{G M_2 \rho}{\left[\rho ^2+(z-z_2)^2\right]^{3/2}}.
\label{eq:frho}
\end{equation}
\end{subequations}
Due to the axial symmetry, there is no special force in the polar direction. In order to find the special points, we calculate the regions where the forces are zero. There are four special points:  
\begin{itemize}
\item {\textbf{Points A, C} $(0,\pm z_0)$:} The force at these points is zero, ${\bf F}_{A,C} = {0}$. For $\rho = 0$, we get $F_\rho = 0$ from Eq.~(\ref{eq:frho}). Imposing $F_z = 0$ in Eq.~(\ref{eq:fz}) gives:
\begin{equation}
\frac{G M_1}{(z_{0}-z_1)^2}+\frac{G M_2}{(z_{0}-z_2)^2} = \frac{\Lambda  c^2  z_{0}}{3}.
\label{eq:ZRASAC}
\end{equation}

\item {\textbf{B-D Path} $(\pm\rho_0,0)$} i.e. the diameter of a circle between the points B and D.

The force towards the center of mass is zero, $F_\rho = 0$. For a perfect dipole ($M_1 = M_2$), the force at these points is exactly zero. The surface  is given by:
\begin{equation}
\frac{G M_1}{\left(\rho_0^2+ z_1^2\right)^{3/2}}+\frac{G M_2}{\left(\rho_0^2+ z_2^2\right)^{3/2}} = \frac{\Lambda c^2}{3}.
\label{eq:ZRASBD}
\end{equation}
The force at the surface is:
\begin{equation}
F_z^{\text{(BD)}} = G M_1 z_1 \left(\frac{1}{\left(\rho_0 ^2+z_1^2\right)^{3/2}}-\frac{1}{\left(\rho_0 ^2+ z_2^2\right)^{3/2}}\right),
\end{equation}
where we used Eq.~(\ref{eq:CoM}) for simplification. For the case of a perfect dipole ($M_1 = M_2$), we get $F_z = 0$.
\end{itemize}

\paragraph{Radial acceleration surface} To determine the boundaries of the regions in the $\left(\rho,z\right)$, we use the ZRAS given by Eq~(\ref{eq:zasCon}). For the galaxy pair case, the surface is given by:
\begin{equation}
\frac{\Lambda c^2  r^2}{3}=G M_1  \frac{r^2-z z_1}{\left(r^2+z_1 (z_1-2 z)\right)^{3/2}}+G M_2 \frac{ r^2-z z_2}{\left(r^2+z_2 (z_2-2 z)\right)^{3/2}}.
\label{eq:zeroRadSur}
\end{equation}
Fig.~(\ref{fig:perDip}) shows in dashed black line the solution for this curve. The plot shows that this curve differentiates the gravitational attraction from the global cosmic expansion. 

\begin{figure*}[t!]
\centering
\includegraphics[width=0.48\textwidth]{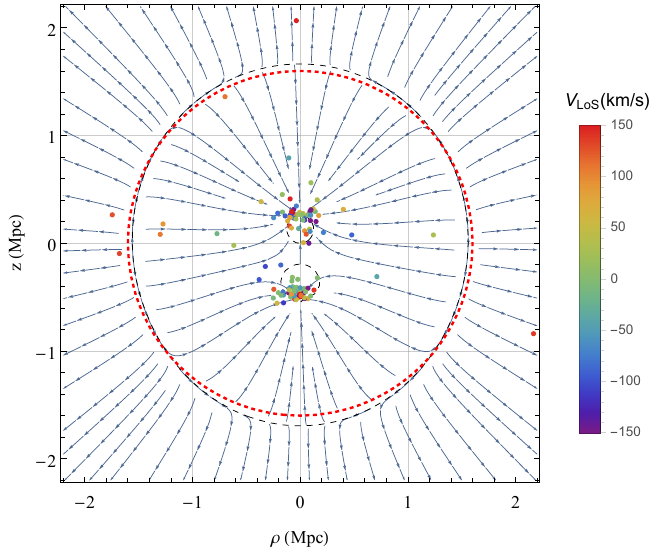} 
\includegraphics[width=0.48\textwidth]{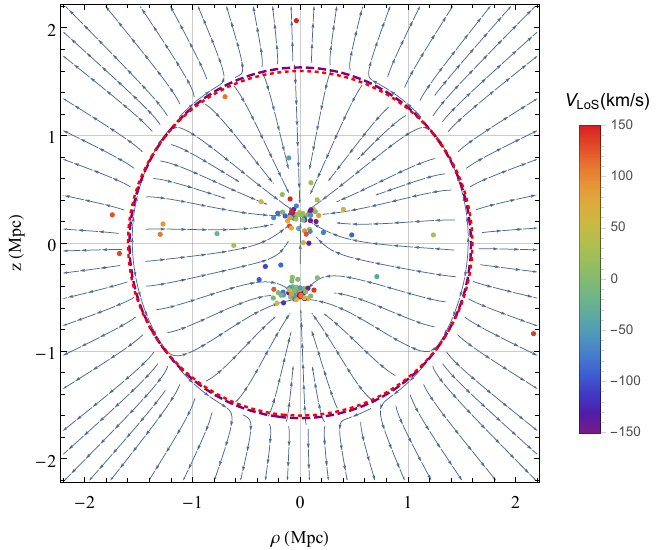} 
\caption{ \it{ Gradient lines (in blue) are for the Local Group of galaxies including the MW (with $M_1 \approx 10^{12} \, M_{\odot}$) and M31 ($M_2 \approx 2 \cdot 10^{12} \, M_{\odot}$) with a distance of $0.77 \, Mpc$, located at $(0,0.77\cdot2/3)$ and $(0,-0.77/3)$, respectively. The center of mass is at the origin of the system of coordinates. The streamlines show the areas for the domination of the Newtonian attraction vs the cosmic repulsion. The dwarf galaxies around the MW and M31 are also added as points with the corresponding colour of the velocity from the LG center of mass. \textbf{Left:} The ZRAS is shown with dashed black lines vs the monopole case in red. \textbf{Right:} The TS is shown in purple vs the monopole case in red. Both definitions for group boundaries are close to the monopole case. The $\bar{r}_{0}$-surface for the monopole case is plotted as the dashed red circle in both panels. }}
\label{fig:streamLinesDipole}  
\end{figure*}

\paragraph{Turnaround surfaces}
Based on the condition for positive kinetic energy, the solution for Eq.~(\ref{eq:con}) gives:
\begin{equation}
\left(\frac{r}{\bar{r}_0}\right)^3=\frac{m_1}{\sqrt{1 - 2 z z_1/r^2+z_1^2/r^2}}+\mathcal{O}\left(1\rightarrow 2\right).
\end{equation}
For the upper and lower points A and C we insert the values $(0,\pm z_0)$ to get:
\begin{equation}
\left(\frac{z_0}{\bar{r}_0}\right)^3=\frac{m_1}{1 - 2 m_2 z_{12}/z_0}+\mathcal{O}\left(1\rightarrow 2\right). 
\label{eq:TSAC}
\end{equation}
For the B-D path:
\begin{equation}
\left(\frac{\rho_0}{\bar{r}_0}\right)^3=\frac{m_1}{\sqrt{1 +m_2^2 z_{12}^2/\rho_0^2}}+\mathcal{O}\left(1\rightarrow 2\right).  
\label{eq:TSBD}
\end{equation}
Fig.~(\ref{fig:perDip}) shows the TS in red. One can see that it gives a good estimate for the boundary of the galaxy group, and most of the dwarf galaxies are inside the surface. The relations between the edges read: $z_0 > \bar{r}_0 > \rho_0$, which shows that this is an extension of the spherically symmetric case.

\subsection{Expansion}
To study the properties of the potential, one can expand the potential around $z_i/r \ll 1$, assuming that the observer is distant from the group. The approximation gives the following~\cite{Penarrubia:2014oda}:
\begin{equation}
\Phi (r) \approx \frac{G M}{r} \left(1 + \frac{\sigma_{\text{Quad}}}{2r^2} \right) + \frac{1}{6}\Lambda c^2 r^2,
\label{eq:expansionPot}
\end{equation}
where:
\begin{eqnarray}
M &=& M_1 + M_2, \quad \\
\nonumber
\sigma_{\text{Dip}} &\equiv& m_1 z_1 + m_2 z_2 = 0, \\
\nonumber
\sigma_{\text{Quad}} &\equiv&\frac{m_1}{2} z_1^2 + \frac{m_2}{2} z_2^2.
\end{eqnarray}
Since we work in CoM coordinates, i.e. CoM is at the origin (Eq.~\ref{eq:CoM}), the dipole of the system is zero~\cite{Penarrubia:2015hqa}. The dipole, octupole, and all multipoles with odd powers in $r^i$ defined in the center of mass of any system should be equal to zero. In order to quantify the contribution of the quadrupole, we can test the dimensionless quantity
\begin{equation}
\tilde{\sigma}_{\text{Quad}} =  \frac{m_1 z_1^2 + m_2 z_2^2}{2 \bar{r}_{0}^2}.
\end{equation}
The condition for the validity of the expansion is $\tilde{\sigma}_{\text{Quad}} \ll 1$. The theoretical consideration for the validation of this expansion is shown in the next section.

\section{Mapping the Close-by Groups}
\label{sec:localUniverse}


\begin{table}[t!]
    \centering
    \begin{tabular}{|c|c|c|}
\hline\hline
     Quantity &   Value   &  Reference   \\
\hline\hline
Separation &   $0.77 \pm 0.04\, Mpc$   &  \cite{vanderMarel:2012xp}   \\
\hline
MW Mass &   $[0.9,1.3 ] \cdot 10^{12}\,M_{\odot}$    &  \cite{Wang:2019ubx}   \\
\hline
M31 Mass &    $[1.0,2.0] \cdot 10^{12}\,M_{\odot}$  &  \cite{Sawala:2022ayk}   \\
\hline\hline
    \end{tabular}
    \caption{\it{The Local Group data properties that taken as a prior in this paper.}}
    \label{tab:lg}

\centering
    \begin{tabular}{|c|c|c|}
\hline\hline
     Quantity &   Value (Mpc)  &  Equation   \\
\hline\hline
$\bar{r}_0$ & $1.48 \pm 0.01 $ & (\ref{eq:r0def}) \\
\hline\hline
ZRAS A,C & $\left( \pm 1.43 \pm 0.01,0 \right)$ & (\ref{eq:ZRASAC}) \\
\hline
B-D path & $\left(0 , \pm 1.58 \pm 0.01 \right)$ & (\ref{eq:ZRASBD}) \\
\hline\hline
TS A,C & $\left( \pm 1.46 \pm 0.01, 0 \right)$ & (\ref{eq:TSAC}) \\
\hline
B-D path & $\left(0,\pm 1.84 \pm 0.02\right)$ &  (\ref{eq:TSBD}) \\
\hline\hline
    \end{tabular}
    \caption{\it{The properties of the LG surfaces for the different surfaces.}}
    \label{tab:lgpost}
\end{table}

\paragraph{LG scale} -- The Local Group (LG) of galaxies is a simple example of a galaxy pair surrounded by dark matter and dwarf galaxies~\cite{1999A&ARv...9..273V}. The two main massive areas are the two giant galaxies, the Milky Way (MW) and Andromeda (M31)~\cite{Sawala:2022ayk}. Fig.~(\ref{fig:streamLinesDipole}) shows the gradient lines (in purple) for the Local Group of galaxies including the MW and M31 with the separation of $0.77\,\text{Mpc}$. The center of mass is at the origin of the coordinate system. The streamlines show the areas for the domination of the Newtonian attraction vs the cosmic repulsion. The intersection of the ZRAS and the TS with the plane is presented by the dashed black lines and the dashed red lines, respectively.

Table~(\ref{tab:lg}) shows the data we consider for the LG of galaxies to estimate the characteristic surfaces. Table~(\ref{tab:lgpost}) shows the different values for the zero acceleration surface (monopole), the critical points ZRAS and TS. We can see from the figure that the surfaces is non-spherical, where one edge is larger than $\bar{r}_0$ and the other edge is smaller than $\bar{r}_0$. The intersection of this surface with the symmetry plane (as in Fig.~(\ref{fig:streamLinesDipole})) is represented by a slightly asymmetric oval that lies between two circles, where $z_{1}/\bar{r}_0 \approx 33 \%$ and $z_{2}/\bar{r}_0 \approx 20 \%,$ are the partial corrections for the galaxy locations over the monopole ZRAS, with an error less than one percent. For the LG, the corresponding quadrupole correction is $\tilde{\sigma}_{\text{Quad}}^{(\text{LG})} < 3.2 \%$. Therefore, treating the quadrupole as a perturbation over the monopole and Dark Energy parts is a valid assumption.

\begin{figure*}[t!]
\centering
\includegraphics[width=0.8\textwidth]{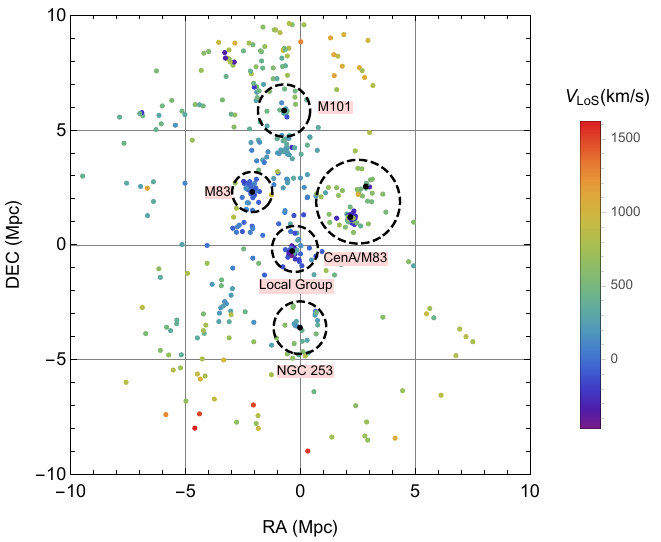} 
\caption{ \it{Two projections of the Local Group (up to 10 Mpc) and their immediate vicinity. The dashed curves show the gravitational bond (with dark energy) for massive galaxies. The figure includes the Local Group, the Cen-A/M83 complex, the M81-M82 group, and the NGC 253 Group. The colours of the dwarf galaxies around show the line of sight velocities as taken from Cosmic Flow~\cite{Hoffman:2023pac}.}}
\label{fig:LocalGroupMap}  
\end{figure*}

\paragraph{Nearby galaxies} -- The formulation applies for different systems in the local Universe~\cite{Karachentsev:2002ex,Karachentsev:2002wf}. In Ref. \cite{Chernin:2015nga} are selected the giant galaxies up to $10\, \text{Mpc}$, and their motion is evaluated. Based on the selection of~\cite{Chernin:2015nga}, we model the local Universe mapping, by assuming the dwarf galaxies are serving as test particles. Fig.~(\ref{fig:LocalGroupMap}) shows the projections of the local Universe (up to 10 Mpc) and their immediate vicinity. The dashed curves show the gravitational bond (with dark energy) for massive galaxies. The colours of the dwarf galaxies around shows the velocity with respect to the CMB frame as taken from Cosmic Flow~\cite{Hoffman:2023pac}. 

\textbf{One Galaxy} -- The simplest case is one dominant giant galaxy and some dwarf galaxies around. The NGC 253 system and the IC-342 group and M101 galaxies seem to satisfies this condition. In this case, the spherical symmetry is a good approximation, unless there is a distribution of dark matter around the system that breaks this symmetry. Therefore, around these giant galaxies the TS looks as a perfect circle, despite that the dark matter distribution could give additional quadrupole corrections. 

\textbf{Galaxy Pair} -- The LG has two main components, the MW and M31. In fact, this binary structure has been suggested as a possible source for the ejection of dwarf galaxies. However, the zero-gravity surface calculated for the LG is almost spherically symmetric at the relevant distances. The \textbf{CenA \& M83 Complex} system seems to be dominated by two giant galaxies: Centaurus A (NGC 5128) and M83 (NGC 5236), where the other galaxies are dwarf galaxies around these. Another pair is the \textbf{M81-M82 Complex}, where the M81 group is a galaxy group in the constellations Ursa Major and Camelopardalis. For these systems, the TS gives an egg shape as discussed in the LG, but since the quadrupole correction is partially small, it looks closer to a circle shape.

\section{Multipole expansion}
\label{sec:mul}
The two examples for one-dominant and two-dominant galaxies are simple cases for a general configuration of galaxies. The gravitational potential $\Phi(r)$ inside the group or cluster is the solution of the Poisson equation: $\Delta \Phi = 4 \pi G \rho - \Lambda c^2$. For a static case, the integration of this equation yields the following:
\begin{equation}
\Phi(r) = G \int \frac{\rho (r')}{|{\bf r} - {\bf r}'|} d^3 r' + \frac{\Lambda 
 c^2 r^2}{6}.
\end{equation}
One can simplify the potential for large distances via the known expansion~\cite{Chaichian:2016owt}:
\begin{equation}
\frac{1}{|{\bf r} - {\bf r}'|} \approx \frac{1}{r} \left( 1 - \frac{\hat{\bf r} \cdot {\bf r}'}{r} + \frac{r'^2 - 3 (\hat{\bf r}\cdot {\bf r}')^2}{2r^2}\right),
\end{equation}
where we take the expansion around the CoM in the origin. The expansion gives the potential as
\begin{equation}
\Phi \approx \frac{G M}{r} \left(1 + \frac{\sigma_{\text{Quad}} (\theta,\phi)}{2r^2} \right) + \frac{1}{6}\Lambda c^2 r^2,
\end{equation}
similarly to Eq.~(\ref{eq:expansionPot}). The mass, the dipole moment, and the quadrupole tensor are:
\begin{eqnarray}
&& M(r) = \int \rho(r') d^3r', \quad {\bf p} = \int \rho(r') \left(\hat{\bf r} - {\bf r}' \right) d^3r' = 0, \nonumber \\
&& Q_{ij} = \int \rho(r') \left[ r'^2\delta_{ij} - 3 (r'_{i} r'_{j})^2 \right] d^3r'.
\end{eqnarray}
The expansion is around the CoM in the origin, therefore ${\bf p} = 0$ \footnote{Notice that the dipole, octupole and and all the multipoles with odd powers in ${\bf r}_i$ defined in the centre of mass  system (or defined in any other system  as odd  powers in ${\bf r}_i-{\bf r}_\text{CoM}$) of any system should be equal to zero, due to the parity invariance of gravitational theories.}. The normalization by the total mass gives:
\begin{equation}
\sigma_{\text{Quad}} (\theta,\phi) = \frac{1}{M}\sum_{ij} Q_{ij} \hat{r}^i\hat{r}^j,
\end{equation}
in terms of the basis components of the quadrupole tensor. Normalizing again by the zero acceleration surface ($\bar{r}_0$) for the monopole case, we obtain a dimensionless parameter that quantifies the validity condition for multipole expansion in our formalism: $\tilde{\sigma}_\text{Quad} = \sigma_\text{Quad}/\bar{r}_0^2 \ll 1 $.

To determine ZRAS, we implement Eq.~(\ref{eq:zasCon}) into the expanded potential, obtaining:
\begin{equation}
\frac{ G  M } {r} \left(1 +\frac{3 \sigma_{\text{Quad}} - \phi \partial_\phi \sigma_{\text{Quad}} - \theta \partial_\theta \sigma_{\text{Quad}} }{2 r^2}\right) = \frac{1}{3} \Lambda c^2  r^2.
\label{eq:ZRASgen}
\end{equation}
One can see that the solution for ZRAS gives the monopole solution for $\sigma_{\text{Quad}} = 0$, via the parameterization 
\begin{equation}
r_{\text{ZRAS}} \approx \bar{r}_0 \left( 1 + \frac{1}{6}A(\theta,\phi) \right),
\end{equation}
where $A$ is a function of the angular variables. Inserting the solution to Eq.~(\ref{eq:ZRASgen}) gives:
\begin{equation}
A + \phi \partial_\phi \tilde{\sigma}_{\text{Quad}} +\theta \partial_\theta \tilde{\sigma}_{\text{Quad}} = 3 \tilde{\sigma}_{\text{Quad}}.
\end{equation}
We can see that the ZRAS is modified due to the geometry of the structure. The TS could be determined in the same way by implementing the condition~(\ref{eq:con}) that gives:
\begin{equation}
G M \left(1 + \frac{\sigma_{\text{Quad}}}{2r_{\text{ts}}^2} \right) = \frac{\Lambda c^2}{3} r_{\text{ts}}^3,
\end{equation}
where $r_{ts}$ is the TS radius. The approximate solution gives:
\begin{equation}
r_{\text{ts}} \approx \bar{r}_0 \left(1 + \frac{1}{6} \tilde{\sigma}_{\text{Quad}}(\theta,\phi) \right) .
\end{equation}
This solution shows that up to a certain correction, the radial solution is valid, with angular dependent solution that comes from the structure of the group of galaxies. Since these groups are in the late Universe, they nearly reached virialization and therefore the solutions seem to be valid for these systems. For the cases of giant galaxy pairs, as we saw, there is a slight quadrupole contribution. Despite that these pairs seem to obey an infall model with the two-body problem, the virialization with other dwarf galaxies could be applied. Notice the elegant correspondence between the solutions for the ZRAS and the TS values where there is an angular modification to $\bar{r}_0$ in both cases. The modification depends on the structure, but both solutions are perturbations to the monopole case.

\section{Estimated Local Group Masses with $\Lambda$}
\label{sec:massesLambda}
To show the importance of $\Lambda$ in a galaxy group, we estimate its contribution to the masses of these groups. We modeled the groups to be dominating by one or two
giant galaxies $(M1,M2)$ and taking $N$ remote galaxies of
similar average mass $m$. Based on our derivation, we
use their detailed locations and velocities from the CoM
for these two galaxies. From Eq.~(\ref{eq:vir}), the kinetic energy average is:
\begin{equation} \label{eq:ke}
\left\langle K  \right\rangle = \frac{3}{2} m N \sigma_v^2,
\end{equation}
where $\sigma_v$ is the one-dimensional velocity dispersion of the system, and the factor of three accounts for the total dispersion under the assumption of isotropy.

Since the measurement is along the line-of-sight velocities, one has to subtract the Hubble flow via: $\mathbf{v}_{\text{LoS}} = \mathbf{v}_i + H_0 \mathbf{r}_i$, where $\mathbf{v}_{\text{LoS}}$ is the measured line-of-sight velocity and $r_i$ is the vector distance to the individual galaxy, and $v_i $ is the line-of-sight component of the peculiar motion. We assume that self-gravity of the outer galaxy groups is negligible compared to the external influence inside the group. The average gravitational potential is:
\begin{equation}
 \left\langle U_G  \right\rangle = - G m N \left( \frac{M_1}{r_{1G}} + \frac{M_2}{r_{2G}} \right),
\end{equation}
where we define the length $r_{G} = N^2 \sum_{i} 1/r_{i0}$ that runs between the dwarf galaxies with respect to the giant ones. The average dark energy potential is:
\begin{equation}
\left\langle U_\Lambda \right\rangle  = \frac{1}{2} m N \Omega_\Lambda H_0^2 \sigma_r^2. 
\end{equation}
Altogether, the virial theorem yields: 
\begin{equation}
3 \sigma_v^2 + \frac{1}{2} \Omega_\Lambda H_0^2  \sigma_r^2 = G M \left( \frac{\gamma}{r_{1G}} +  \frac{1 - \gamma}{r_{2G}} \right),
\end{equation}
where $\gamma = M_1/M$. For a system with only one giant galaxy, $M_2 = 0$, and then we consider the center to be around $M_1$. For a given ratio of the giants, it is possible to infer the total mass from the dispersions of velocities and the distances. For a comparison~\cite{Diaz:2014kqa,Hartl:2021aio} uses $\gamma = M_{31}/M_{MW} = 2.3$ and obtains $\left(2.5 \pm 0.4\right) \cdot 10^{12}\, M_{\odot}$. The contribution of $\Lambda$ increases the predicted mass by the ratio:
\begin{equation}
M_{\text{VT} + \Lambda} = M_{\text{VT}} \left[1 + \frac{\Omega_\Lambda}{6} \left(\frac{H_0 \sigma_r}{\sigma_v}\right)^2 \right],
\end{equation}
where $M_{VT}$ is the estimated mass considering the Newtonian interaction and  $M_{VT + \Lambda}$ is the estimated mass with $\Lambda$. 

\begin{figure}[t!]
\centering
\includegraphics[width=0.43\textwidth]{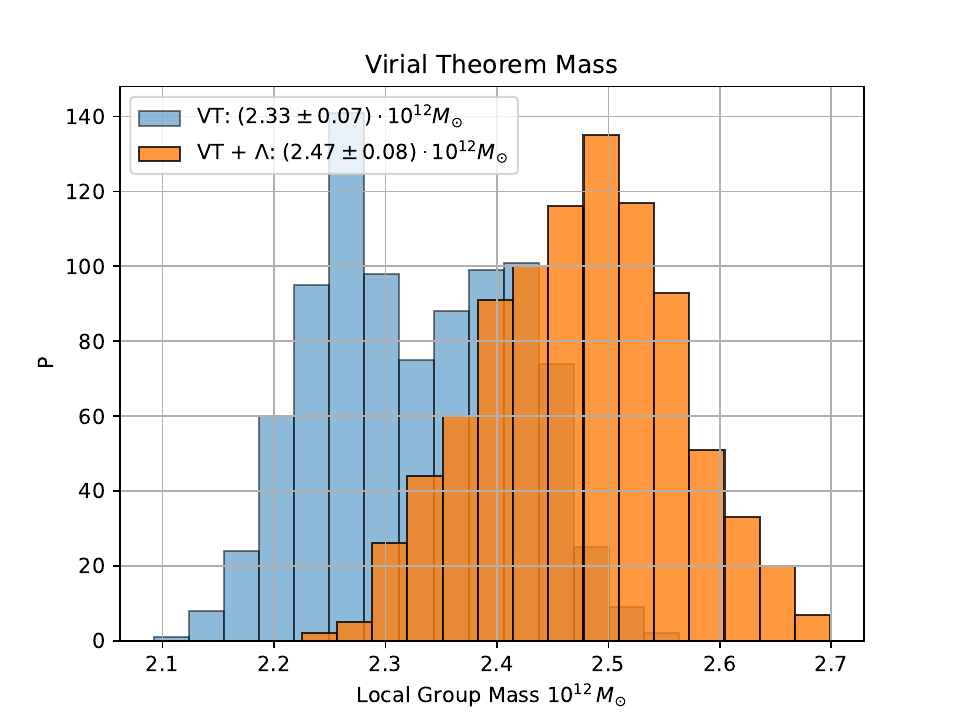} 
\caption{\it{The posterior distribution for the LG total mass using the virial theorem with and without $\Lambda$. $\Lambda $ creates a repulsion force and to compensate for this additional interaction, the virialized mass of the system has to be higher.}}
\label{fig:perLGVir}  
\end{figure}

We assume a Gaussian prior on: $H_0 = 67.4\pm0.5 \, km/s/Mpc$, $\Omega_\Lambda = 0.685 \pm 0.007$ and 5\% error on the velocities. The enclosed mass of the LG is estimated to be: $M_{LG}^{(VT)} = \left( 2.33 \pm 0.07\right) \cdot 10^{12} M_{\odot}$. With $\Lambda $, the mass increases to $M_{LG}^{(VT + \Lambda)} = \left( 2.47 \pm 0.08\right) \cdot 10^{12} M_{\odot}$. Fig.~(\ref{fig:perLGVir}) shows the posterior distribution for the estimated masses with and without $\Lambda$. The mass is shifted by $\sim 5-10\%$ which is within the errors of the estimated masses. However, with future measurements, especially around the turnaround surface, it will be possible to increase the precision and the effect of $\Lambda$ will be seen as a stronger one. 

\section{Conclusions}
\label{sec:dis}
The present work extends the definitions of characteristic surfaces for the galaxy groups to include the effect of dark energy. In this general framework, the interplay between the attraction of matter and dark-matter distribution and the repulsion force of dark energy gives a natural definition for the local groups. Within this approach, we model the dwarf galaxies as test particles around the giant galaxies. In earlier studies, galaxies were grouped on the basis of the observed distribution of the considered galaxies, assuming only Newtonian interaction, and subsequently found the masses enclosed in the groups. In the present work, however, we emphasize the importance of dark energy inside the galaxy groups and in addition we extend the definitions of the boundaries of galaxy groups by including the local effect of the cosmological constant $\Lambda$.

For the monopole case, where there is one giant galaxy, there is a clear radius where the total force is zero, namely the zero gravity radius. By considering the system to obey the virial theorem, one can shows that this radius is also the upper radius for the virialized system. For a galaxy pair, the total acceleration for the monopole case is generalized to the Zero Radial Acceleration Surface (ZRAS) that obeys Eq.~(\ref{eq:zasCon}). The turnaround surface is obtained by setting the kinetic energy of dwarf galaxies to be positive. In this case, a potential surface that obeys Eq.~(\ref{eq:con}) is obtained and gives a boundary for the group.

The cosmological constant has a small but measurable effect on the motion of the Milky Way (MW) and Andromeda (M31) galaxies, which are the two largest members of the Local Group. The zero acceleration surface of the LG has a radius that is about $1.5 \, \text{Mpc}$, i.e. about twice the separation between these galaxies. However, this surface is only determined by assuming spherically symmetric behavior around the total mass of the LG. In this paper, we extend the definitions for the ZRAS and TS for galaxy pairs, and we determine these for the LG. We show that there is a small correction due to the quadruple contribution.

With these two simple examples, we map the galaxies in the local Universe based on the data from CosmicFlow. We consider different giant galaxies and galaxy pairs, and we show that the close-by groups are separated. The dynamics of these groups can be determined by the outside and inside the boundary: Inside the boundary, the system will become closer to a spherically symmetric solution in the future (the pair will fall into the center of mass) and the groups will continue to distance themselves apart from each other. We generalize these solution using multipole expansion, and we show that choosing the correct center of mass of system yields an elegant general description: The system could be determine by the monopole and the quadrupole moments (since the dipole is zero) and the boundaries could be determine by the interplay of gravitational attraction and dark energy.

By invoking the powerful Virial Theorem \footnote{The Virial Theorem has previously been used in astrophysics in~\cite{Milgrom:2013fva} for the modified gravitational theory, MOND; see also~\cite{Milgrom:1994ge} for a general derivation of the theorem for action-governed theories, and as well for some historical references therein on many applications of the theorem.}, we estimate the LG mass, assuming that dwarf galaxies around MW and M31 have a very small mass. We show that the estimated LG mass increases by 5 to 10\% to be $\left( 2.47 \pm 0.08\right) \cdot 10^{12} M_{\odot}$.

The result is consistent with other independent measurements of the LG mass. Specifically, it aligns with values reported in the literature, which also suggest $\sim[2,3] \cdot 10^{12} M_\odot$ based on fitting the motions of dwarf galaxies to the Hubble flow~\citep{Peirani:2005ti,Peirani:2008qs,Karachentsev:2008st,Penarrubia:2014oda,Teerikorpi:2010zz,DelPopolo:2022sev}. Additionally, considering the Milky Way and Andromeda (M31) as a binary system with the Large Magellanic Cloud acting as a perturbation~\cite{Benisty:2024lsz}, a mass of $\left(2.3 \pm 0.7\right) \cdot 10^{12}M_{\odot}$ has been derived.

We compared our total mass estimates with the sum of the individual galaxy masses. Various studies provide values for the separate masses of each galaxy: $M_{MW} \in [0.9,1.3] \cdot 10^{12} M_{\odot}$ and $M_{31} \in [1.,2.] \cdot 10^{12}, M_{\odot}$ \citep{Wang:2019ubx,Sawala:2022ayk}. The total mass of both galaxies falls within the range $M_{MW} + M_{31} \in [1.9,3.3] \cdot 10^{12} M_{\odot}$. The LG mass should at least be within this range, though higher values might arise due to the presence of undetected dark matter around the galaxies. The total mass estimated from the virial theorem (including $\Lambda$) is consistent with this range, suggesting that additional dark matter between the galaxies may not be necessary. However, this remains an open question, as some studies, such as Ref.~\cite{Sawala:2023sec}, propose additional dark matter between the galaxies based on simulations.

The whole work shows the importance of invoking the effect of $\Lambda$, which has previously been ignored in studies of the local universe. Moreover, we show that the definition of groups naturally comes from the interplay between the effect of $\Lambda$ with other constituents of matter and their interaction, and therefore should be considered in future studies, even within close-by scales.

\acknowledgments
We are grateful to Noam Libeskind, Mordehai Milgrom, Paolo Salucci, Horst Stocker, and in particular Jenny Wagner for the illuminating discussions. DB is supported by a Minerva Fellowship of the Minerva Stiftung Gesellschaft fuer die Forschung mbH.

\bibliographystyle{apsrev4-1}
\bibliography{ref.bib}

\end{document}